\documentclass[journal,10pt]{IEEEtran}

\usepackage[cmex10]{amsmath}
\usepackage[utf8]{inputenc}
\usepackage[pdftex]{graphicx}
\usepackage{epstopdf}
\usepackage{array}
\usepackage{floatrow}
\usepackage{caption}
\usepackage{hyperref}[hidelinks]
\usepackage{cleveref}
\usepackage{subcaption}
\usepackage{url}
\usepackage{balance}
\usepackage{enumerate} 
\usepackage{amsbsy}
\usepackage[normalem]{ulem}
\usepackage{textcomp}
\usepackage{siunitx}
\usepackage{balance}
\usepackage{placeins}
\usepackage{tabularx, ragged2e}
\usepackage{listings}
\usepackage[dvipsnames]{xcolor}
\usepackage[T1]{fontenc}
\usepackage{helvet}
\usepackage{comment}

\def\BibTeX{{\rm B\kern-.05em{\sc i\kern-.025em b}\kern-.08em
    T\kern-.1667em\lower.7ex\hbox{E}\kern-.125emX}}
    
\usepackage{xspace}

\DeclareTextFontCommand{\textcomputer}{\fontfamily{cmr}\selectfont}

\begin{document}

\title{Measurement-Based Performance Evaluation of SmartRSUs with Heterogeneous Antenna Architectures for V2X Communications}

\author{Marco~Savarese, Gaetano~Orazio~Cauchi, Salvatore~Iandolo, Antonio~Solida,
        Martin~Klapez, Maurizio~Casoni, Micaela~Verucchi, Enrico~Vincenzi, Ignacio~Sanudo~Olmedo, Marko~Bertogna, and~Carlo~Augusto~Grazia
\thanks{M. Savarese, G. O. Cauchi, S. Iandolo, A. Solida, M. Klapez, M. Casoni, and C.A. Grazia are with the Department of Engineering \emph{Enzo Ferrari}, University of Modena and Reggio Emilia, via Pietro Vivarelli, 10, 41125, Modena, Italy
(e-mail: \{\emph{name}.\emph{surname}\}@unimore.it).
\\
M. Verucchi, E. Vincenzi, I. Sanudo Olmedo, and M. Bertogna are with Hipert S.r.L., via della Scienza, Modena, Italy (e-mail: \{\emph{name}.\emph{surname}\}@hipert.it).
This study was carried out within the MOST – Sustainable Mobility National Research Center and received funding from the European Union Next-GenerationEU (PIANO NAZIONALE DI RIPRESA E RESILIENZA (PNRR) – MISSIONE 4 COMPONENTE 2, INVESTIMENTO 1.4 – D.D. 1033 17/06/2022, CN00000023). This manuscript reflects only the authors’ views and opinions, neither the European Union nor the European Commission can be considered responsible for them.}
}

\maketitle

\begin{abstract}

This paper presents a measurement-based performance evaluation of two custom Smart Roadside Units (SmartRSUs) featuring different V2X antenna architectures. The first configuration integrates GNSS and communication antennas into an all-in-one rooftop module, whereas the second uses external dual ITS-G5 (IEEE 802.11p) antennas operating at 5.9~GHz and a dedicated GNSS antenna. Both systems are built upon a proprietary On-Board Unit (OBU) platform adapted for infrastructure deployment.

The experimental campaign evaluates key V2X communication metrics, including coverage, received signal strength indicator (RSSI), packet loss, and end-to-end latency in both transmission (OBU-to-infrastructure) and reception (infrastructure-to-OBU) directions. To ensure objective validation, a commercial off-the-shelf V2X Roadside Unit is co-located on the same infrastructure and used as a performance benchmark, providing ground-truth reference measurements under identical environmental conditions through a controlled co-located deployment.

Results highlight the impact of antenna design and placement on communication reliability and latency, revealing trade-offs between integrated and external antenna configurations in real-world deployment scenarios. The findings provide practical insights for the design and optimization of next-generation SmartRSUs in cooperative intelligent transportation systems (C-ITS).

\end{abstract}


\section{Introduction}
\label{sec_intro}

The development of connected and automated driving systems is increasingly shifting from purely vehicle-centric intelligence toward a cooperative paradigm, where smart infrastructure plays a key role in enhancing perception, reliability, and safety. In this context, Vehicle-to-Everything (V2X) communication—particularly based on ITS-G5 (IEEE 802.11p) operating at $5.9~GHz$ remains a fundamental enabler of low-latency, direct information exchange between vehicles and the surrounding environment. Ensuring robust and predictable V2X performance in real-world deployments is therefore a critical requirement for both research validation and pre-deployment testing.

Road-Side Units (RSUs) are central components of this ecosystem, acting as communication hubs between vehicles and infrastructure services. However, their performance is strongly influenced by hardware design choices, especially antenna configuration and placement, which directly affect coverage, signal quality, and communication reliability. Despite their importance, these aspects are often evaluated in isolation or under limited experimental conditions, lacking direct comparison with reference systems in realistic environments.

The Modena Automotive Smart Area (MASA) is a unique large-scale living lab for cooperative and connected mobility, designed to support the experimental validation of V2X technologies in controlled yet realistic urban conditions. Within MASA, a new infrastructure node has been recently deployed, hosting multiple RSUs on the same pole, including both custom-developed and commercial systems. This setup enables direct, fair, and repeatable comparison across different hardware architectures under identical environmental and operational conditions.

In this work, we present a measurement-based performance evaluation of two in-house SmartRSUs, both built on our previously introduced Open DSRC Unit (ODU) platform~\cite{odu}. The ODU is an open, modular hardware/software solution for V2X communication that has been progressively extended from vehicle-mounted On-Board Units (OBUs) to infrastructure-based deployments. As a result, MASA now features a fully integrated ecosystem of custom OBUs and RSUs, enabling end-to-end experimentation with open and reproducible technologies.

The two SmartRSUs analyzed in this paper differ in their antenna architectures: one uses an integrated all-in-one rooftop module that combines GNSS and communication antennas, while the other employs external dual ITS-G5 antennas with a dedicated GNSS unit. Their performance is experimentally evaluated in terms of coverage, received signal strength indicator (RSSI), packet loss, and end-to-end latency, considering both uplink (OBU-to-infrastructure) and downlink (infrastructure-to-OBU) communication. To provide a reliable ground-truth reference, a commercial off-the-shelf RSU is co-located on the same pole and used as a benchmarking system.

Beyond the specific comparison, this work is motivated by the broader vision that future automated and remotely operated driving systems will rely heavily on smart-city infrastructure as “training grounds” for advanced cooperative applications. From this perspective, SmartRSUs that integrate sensing, communication, and computation capabilities can enable innovative services—such as AI-assisted perception systems (\emph{e.g.}, AI-CAM~\cite{aicam})—within an open, replicable framework. The MASA infrastructure, built on open, modular solutions such as the ODU, aims to support scalability and reproducibility across diverse deployment scenarios.

The main contributions of this paper can be summarized as follows:
(i) the design and deployment of two ODU-based SmartRSUs with different antenna configurations;
(ii) a comprehensive, measurement-based comparison of their V2X communication performance;
(iii) a co-located benchmarking methodology enabling fair comparison under identical conditions;
(iv) quantitative insights into the impact of antenna architecture on real-world V2X performance.

The paper is organized as follows:
\Cref{sec_related} overviews related works.
\Cref{sec_masa} describes the MASA living lab; Details on its sensing, networking, and computing architecture will be provided.
\Cref{sec_testbed} details our experimental setup.
\Cref{sec_test_results} presents and analyzes the results, while
\Cref{sec_conclusions} concludes the article.

\section{Related Works}
\label{sec_related}

Few studies have resulted in open-source solutions operating at ITS frequencies. For example, Raviglione \emph{et al}. in~\cite{demo_raviglione} developed a Linux-based, open-source platform for evaluating IEEE 802.11p-compatible Network Interface Cards, following the cost-effective hardware approach in~\cite{7528312}. Their solution relies on APU1D embedded boards equipped with the Unex DHXA-222 WiFi PCIe module and the OpenWrt operating system.

Despite its open-source nature, this approach presents several limitations. IEEE 802.11p support requires manual patching, while the hardware platform is relatively large, expensive, and no longer available, limiting scalability and deployment in emerging scenarios such as VRU applications. The same authors also proposed Oscar in~\cite{oscar_raviglione}, an open-source ETSI-compliant C-ITS stack supporting Cooperative Awareness Messages (CAMs) and Vulnerable Awareness Messages (VAMs), with planned extensions to additional ETSI services. However, its strict compliance with ETSI specifications reduces flexibility, preventing operation without GNSS input and limiting the use of custom messages.

Alternative approaches have explored programmable platforms. Router-based solutions have been adapted to operate as RSUs or OBUs in the $5.9~GHz$ band~\cite{martin1, martin2}, enabling higher transmission power but at the cost of increased size and reduced portability. Similarly, Software Defined Radio (SDR) platforms~\cite{8031977} and Hardware-in-the-Loop (HiL) approaches~\cite{POSTER} provide high flexibility but introduce additional system complexity. Bridging the gap between simulation and real-world experimentation remains a key challenge. As shown in~\cite{almeida-sim-real}, simulation results may significantly deviate from real deployments, particularly in terms of communication range and packet delivery, highlighting the need for measurement-based experimental platforms.

From an application perspective, works such as~\cite{v2video} and~\cite{v2video_old} investigate IEEE 802.11p-based video transmission but often lack sufficient detail on hardware and protocol configurations, thereby limiting reproducibility and validation. To address these limitations, open and modular platforms have been recently proposed. The Open DSRC Unit (ODU)~\cite{odu} is a low-cost solution based on off-the-shelf components that operates as both an OBU and an RSU, enabling reproducible experimentation and rapid prototyping of V2X systems.

Building on such platforms, infrastructure-assisted approaches have been introduced to support advanced cooperative services. The AI-CAM framework~\cite{aicam} leverages roadside sensing to generate ETSI-compliant Cooperative Awareness Messages, extending perception capabilities to non-connected entities.

Finally, real-world living labs such as the Modena Automotive Smart Area (MASA) enable experimental validation of V2X systems under realistic conditions, supporting latency-sensitive applications such as remote driving and cooperative perception~\cite{martin1}. Despite these advances, the impact of RSU hardware design, particularly antenna configuration, on communication performance remains largely unexplored. Most existing works rely on fixed hardware platforms or lack direct comparisons under identical conditions. This gap motivates the measurement-based analysis presented in this work.

\begin{figure}
  \centering
\includegraphics[width=0.99\textwidth]{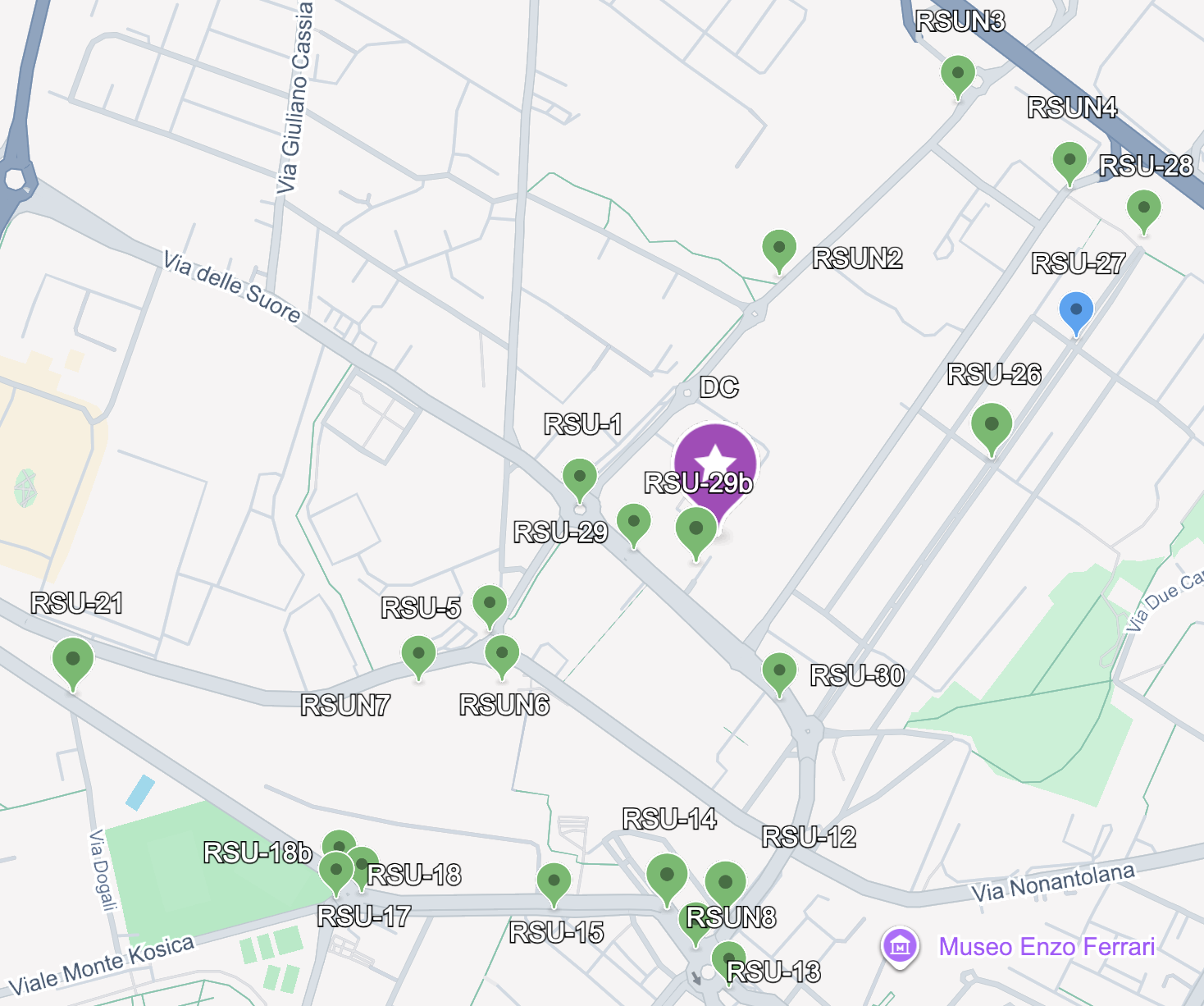}
    \caption{The current deployment state of the MASA living lab: 21 sites (green) equipped with a Smart Camera and an RSU, plus 1 site (blue) used for testing in this paper with 2 SmartRSU and a commercial RSU for reference.}
    \label{fig:masa-live}
\end{figure}

\begin{figure}
  \centering
\includegraphics[width=0.99\textwidth]{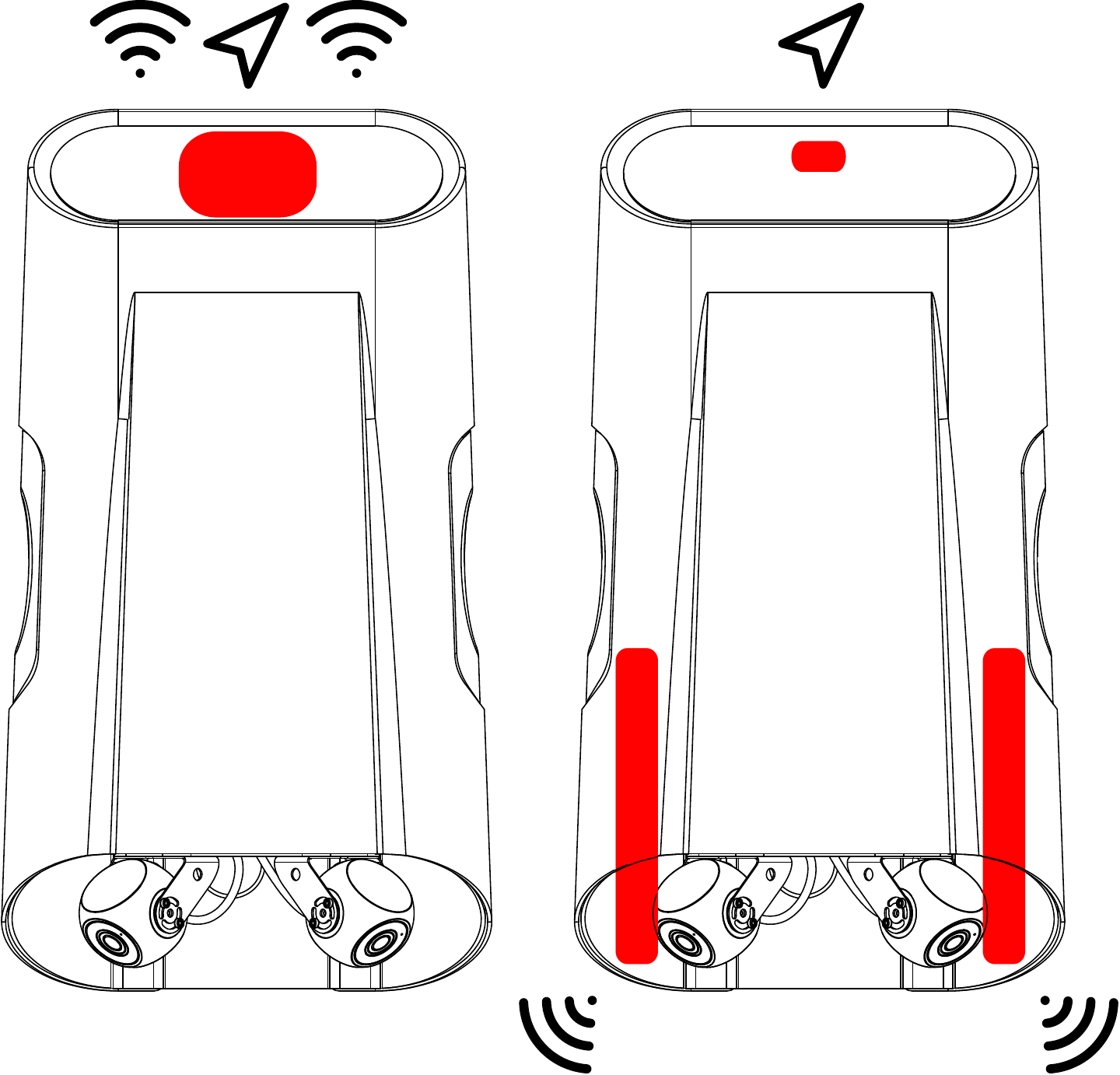}
    \caption{SmartRSU with all-in-one antenna for GNSS and ITS-G5 (left) and SmartRSU with GNSS antenna on top and the two lateral ITS-G5 dipoles (right).}
    \label{fig:smart_rsu}
\end{figure}

\begin{figure}
\subfloat[]{\includegraphics[width=0.56\textwidth]{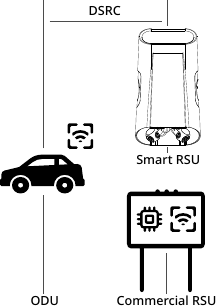}
\label{fig:double_rsua}}
\hfill
\subfloat[]{\includegraphics[width=0.365\textwidth]{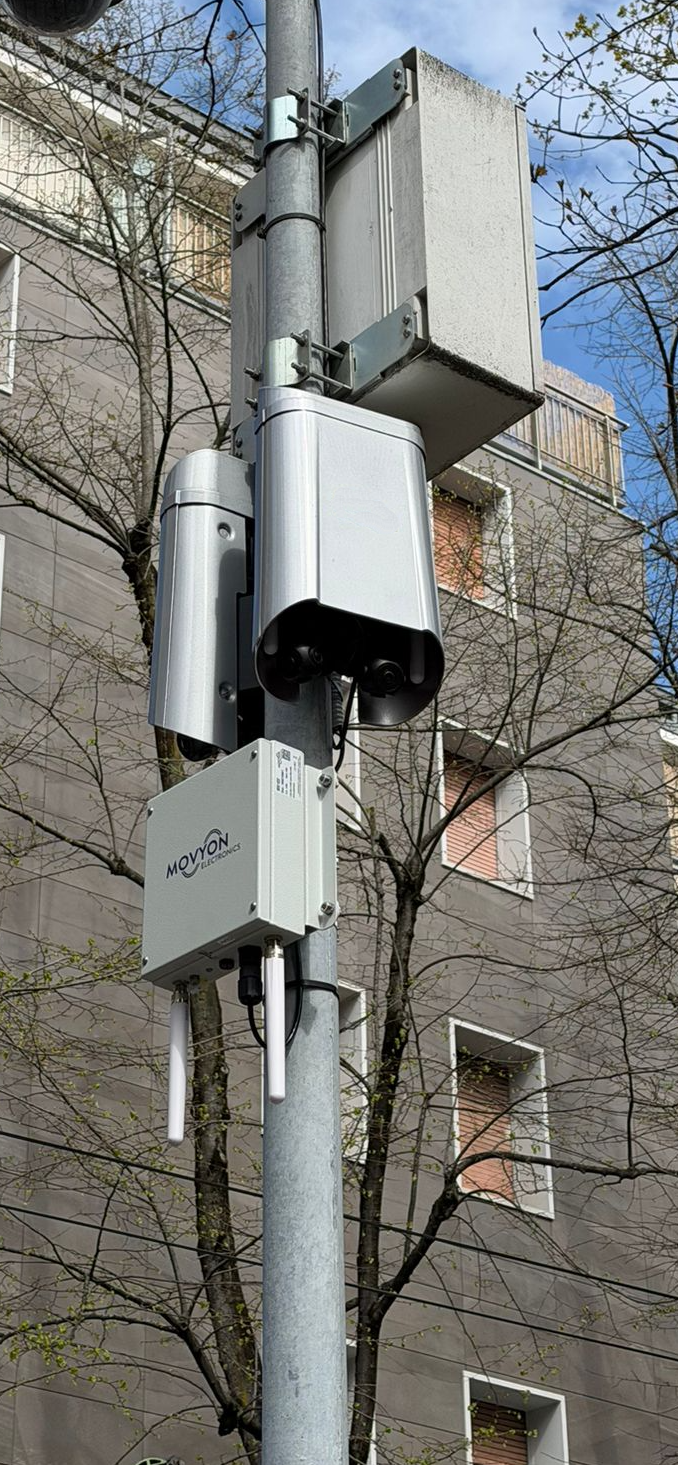}
\label{fig:double_rsub}}
\caption{Experimental setup: (a) logical representation of the communication nodes; (b) physical deployment of the co-located RSUs on the pole.}
\label{fig:double_rsu}
\end{figure}

\section{The MASA Living Lab}
\label{sec_masa}

The Modena Automotive Smart Area (MASA) is an open-air urban living lab designed to support the development, validation, and deployment of connected and cooperative mobility applications under realistic conditions. The infrastructure integrates sensing, communication, and computing capabilities, enabling end-to-end experimentation for V2X systems and latency-sensitive services.

As shown in Fig.~\ref{fig:masa-live}, the current MASA deployment comprises 21 distributed sites, each equipped with a Smart Camera and a Road Side Unit (RSU), interconnected via a high-capacity network backbone. This large-scale infrastructure allows the collection of real-world measurements across heterogeneous urban scenarios, providing a reproducible environment for experimental validation. In addition to the existing sites, a new dedicated node (highlighted in blue in Fig.~\ref{fig:masa-live}) has been deployed for the purpose of this study. This node hosts multiple RSUs co-located on the same pole, including two custom SmartRSUs and one commercial off-the-shelf RSU used as a reference benchmark. This configuration enables a direct and fair comparison of different hardware solutions under identical environmental and operational conditions.

The two SmartRSUs developed in this work are based on the ODU platform and differ in their antenna configuration, as shown in Fig.~\ref{fig:smart_rsu}. The first solution adopts an integrated all-in-one antenna module that combines GNSS and ITS-G5 communications, whereas the second employs a separate design with a rooftop GNSS antenna and two lateral ITS-G5 dipole antennas. These differences allow the investigation of the impact of antenna architecture on V2X communication performance.

Overall, the MASA living lab provides a unique environment for measurement-based evaluation of V2X systems, enabling the validation of communication performance in realistic deployment scenarios and thereby enabling controlled yet realistic evaluation of V2X communication systems.

The selected test site is in a particularly challenging urban area characterized by partial signal obstruction and urban-canyon effects. The presence of buildings and tree-lined roads introduces multipath propagation and non-line-of-sight conditions, which affect not only V2X communication but also GNSS signal quality.

As a result, the positioning information embedded in CAM messages may exhibit noticeable inaccuracies, with deviations of several meters observed during the experimental campaign. This behavior is also evident in the map-based results, where vehicle trajectories appear nonlinear and are affected by localization noise. Such effects are not limited to the experimental platform but are also observed in commercial vehicles operating in the same area. Although GNSS accuracy is not explicitly analyzed in this work, these conditions underscore the importance of infrastructure-assisted services, such as AI-CAM, which can mitigate positioning inaccuracies by leveraging roadside sensing and cooperative perception. This makes the considered scenario particularly relevant for evaluating real-world V2X deployments under non-ideal conditions.
 
\section{Experimental setup}
\label{sec_testbed}

The experimental campaign was conducted within the Modena Automotive Smart Area (MASA), using a dedicated test site with three RSUs co-located on a single pole, as illustrated in Fig.~\ref{fig:double_rsu}. This configuration enables a direct comparison of different RSU implementations under identical environmental conditions.

The testbed includes three RSUs: two custom SmartRSUs developed in-house and one commercial off-the-shelf RSU used as a reference. The SmartRSUs are built by integrating the communication capabilities of the Open DSRC Unit (ODU) into an intelligent camera platform, thus enabling flexible experimentation while maintaining compatibility with standard V2X communication protocols. The two SmartRSUs differ exclusively in their antenna configuration. The first adopts an integrated all-in-one rooftop antenna that combines GNSS and ITS-G5 communications, while the second employs a separate design with a GNSS antenna mounted on top and two external ITS-G5 dipole antennas. The commercial RSU represents the baseline reference, providing a standard-compliant benchmark for performance evaluation.

To ensure a fair comparison, all RSUs are configured to operate under identical communication settings. In particular, each RSU periodically broadcasts ETSI-compliant CAMs at $10~Hz$, corresponding to the upper-bound transmission rate defined by the ETSI standard. Since RSUs are static infrastructure nodes, all transmitted CAMs include a fixed position corresponding to the pole's installation point.

The vehicular side is represented by a moving platform equipped with an ODU configured as an On-Board Unit (OBU). The vehicle transmits CAM messages at $10~Hz$, embedding real-time GNSS position and speed information, and simultaneously receives messages from all RSUs. All packets are captured in monitor mode and stored for post-processing, enabling the extraction of performance metrics, including RSSI, packet loss, and latency.

The experimental campaign comprises five repeated measurement runs, during which the vehicle moves within a radius of approximately $500~m$ of the RSU pole, following the road network available in the MASA area (cf. Fig.~\ref{fig:masa-live}). The test environment includes urban features such as tree-lined roads and partial obstructions, resulting in several non-line-of-sight (NLOS) conditions within the nominal coverage area. This setup allows the evaluation of communication performance under realistic propagation scenarios. In addition to mobility tests, a controlled transmission campaign has been conducted to evaluate latency under static conditions. In this case, the vehicle is positioned close to the RSU pole, and a UDP traffic flow is generated using \texttt{iperf3} at a constant rate of $2~Mbps$. This configuration enables the assessment of transmission latency in a stable, interference-limited scenario without mobility-induced variability.

All RSUs are installed at approximately $3.5~m$, ensuring comparable propagation conditions across all devices. This experimental setup enables a comprehensive, measurement-based comparison of custom and commercial RSUs, isolating the impact of antenna configuration on V2X communication performance.

\section{Test Results}
\label{sec_test_results}

\begin{figure}
  \centering
\includegraphics[width=0.95\textwidth]{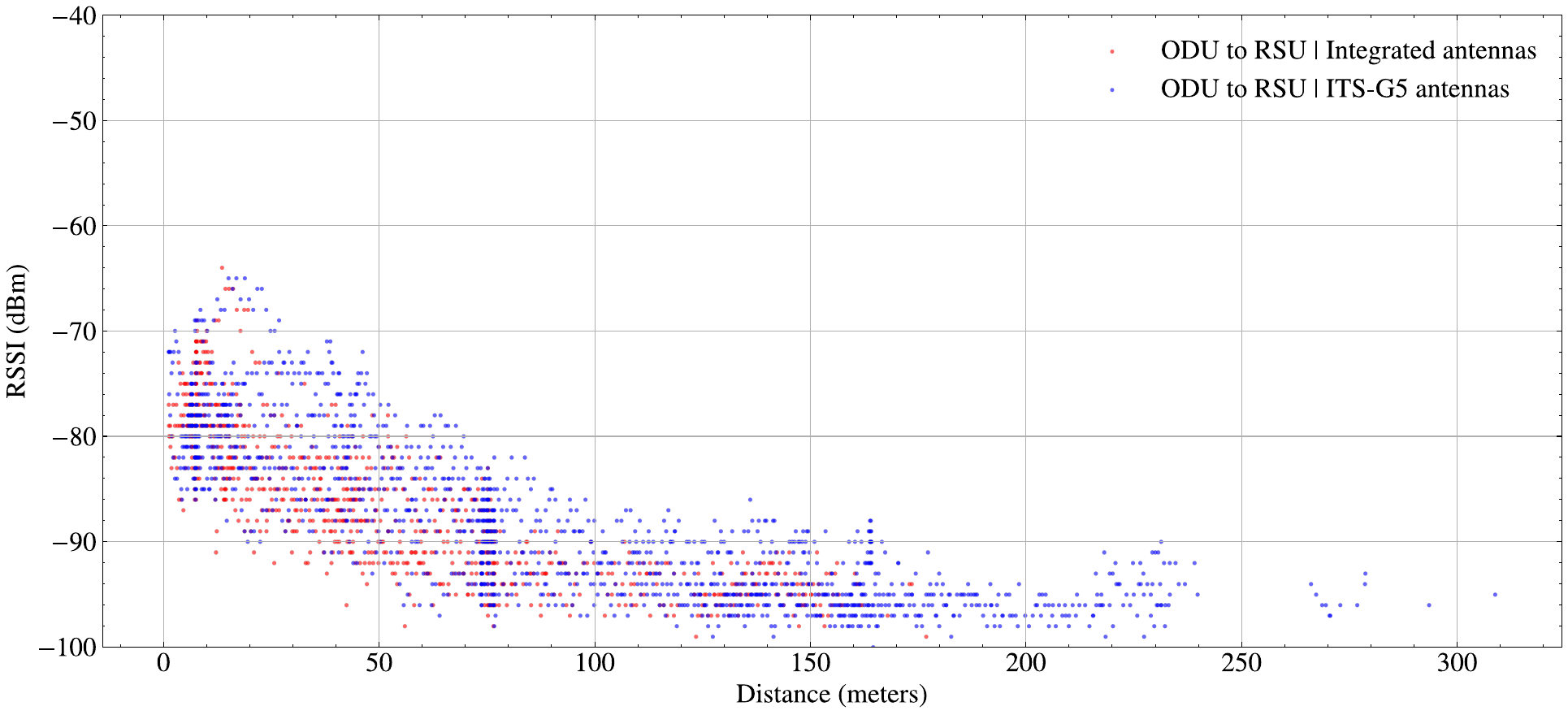}
\caption{Average RSSI received by the SmartRSUs as a function of the distance between the ODU and the RSU pole.}
\label{fig:odu_to_rsu_rssi}
\end{figure}

\begin{figure}
    \subfloat[SmartRSU with integrated antennas.]{\includegraphics[width=0.45\linewidth]{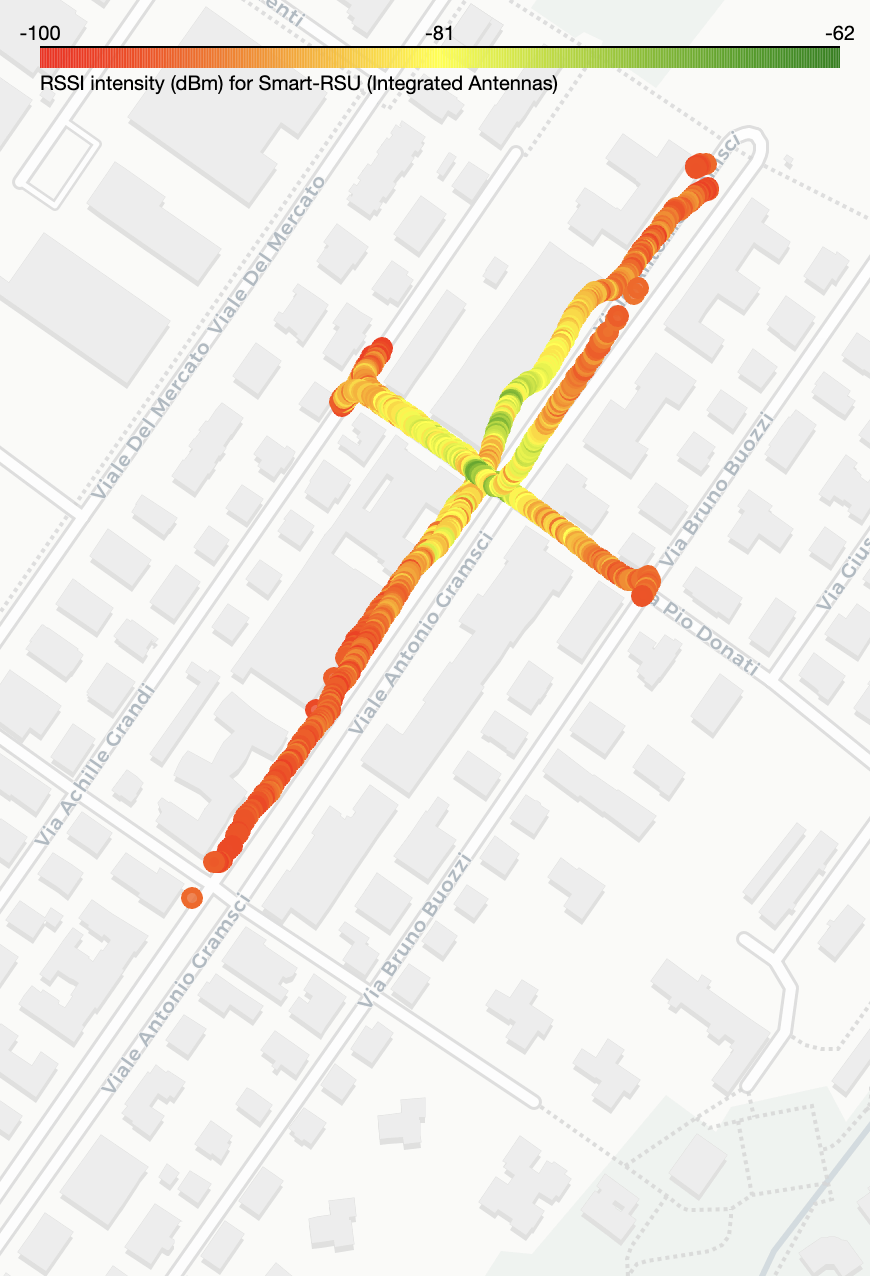}}
    \hfill
    \subfloat[SmartRSU with ITS-G5 antennas.]{\includegraphics[width=0.45\linewidth]{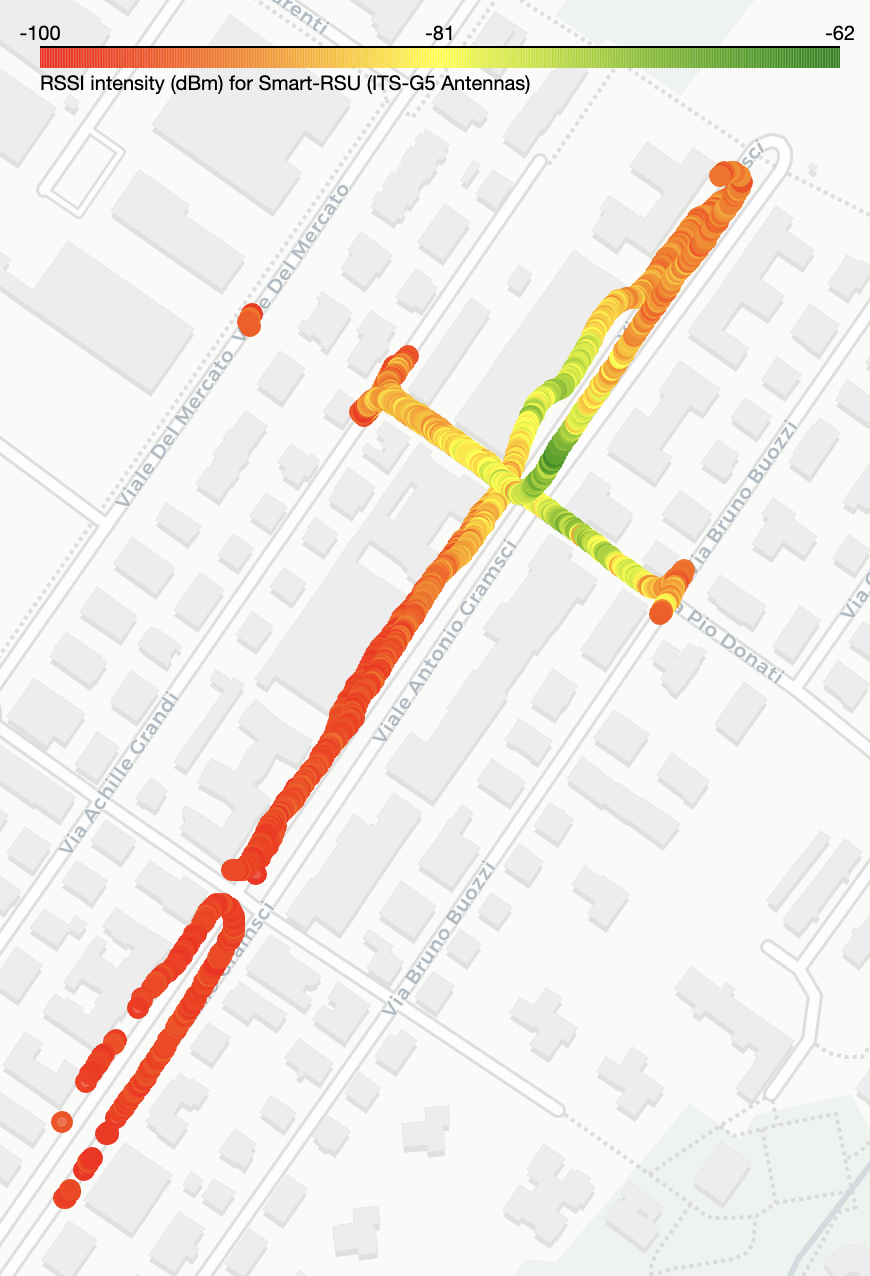}}
    \label{fig:rssi166}
    \caption{Spatial distribution of RSSI received by each of the SmartRSUs.}
    \label{fig:maprssi}

\end{figure}

\begin{figure*}
\subfloat[]{\includegraphics[width=0.32\textwidth]{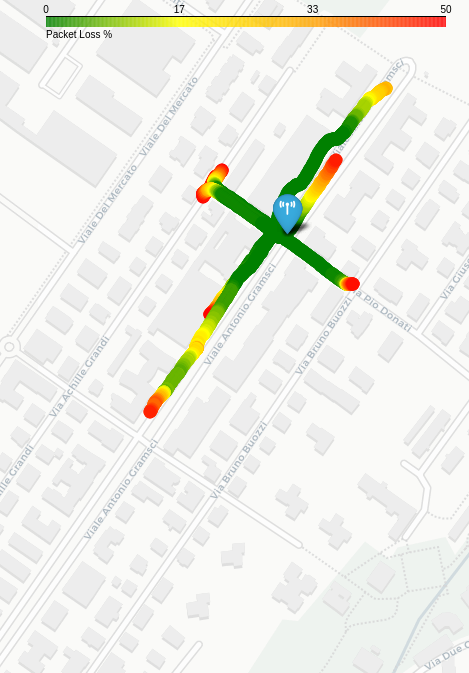}
\label{fig:maplossint}}
\hfill
\subfloat[]{\includegraphics[width=0.32\textwidth]{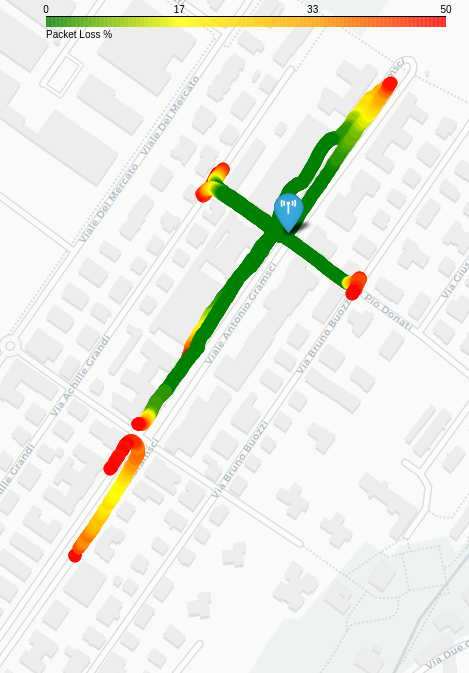}
\label{fig:maplossext}}
\hfill
\subfloat[]{\includegraphics[width=0.32\textwidth]{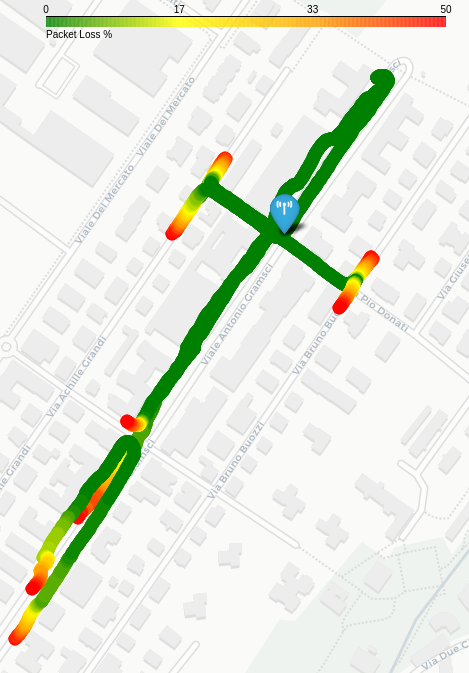}
\label{fig:maplosscomm}}
\caption{Spatial distribution of packet loss for all RSUs: (a) SmartRSU with integrated antennas; (b) SmartRSU with external ITS-G5 antennas; (c) commercial RSU.}
\label{fig:maploss}
\end{figure*}

\begin{figure}
  \centering
\includegraphics[width=0.95\textwidth]{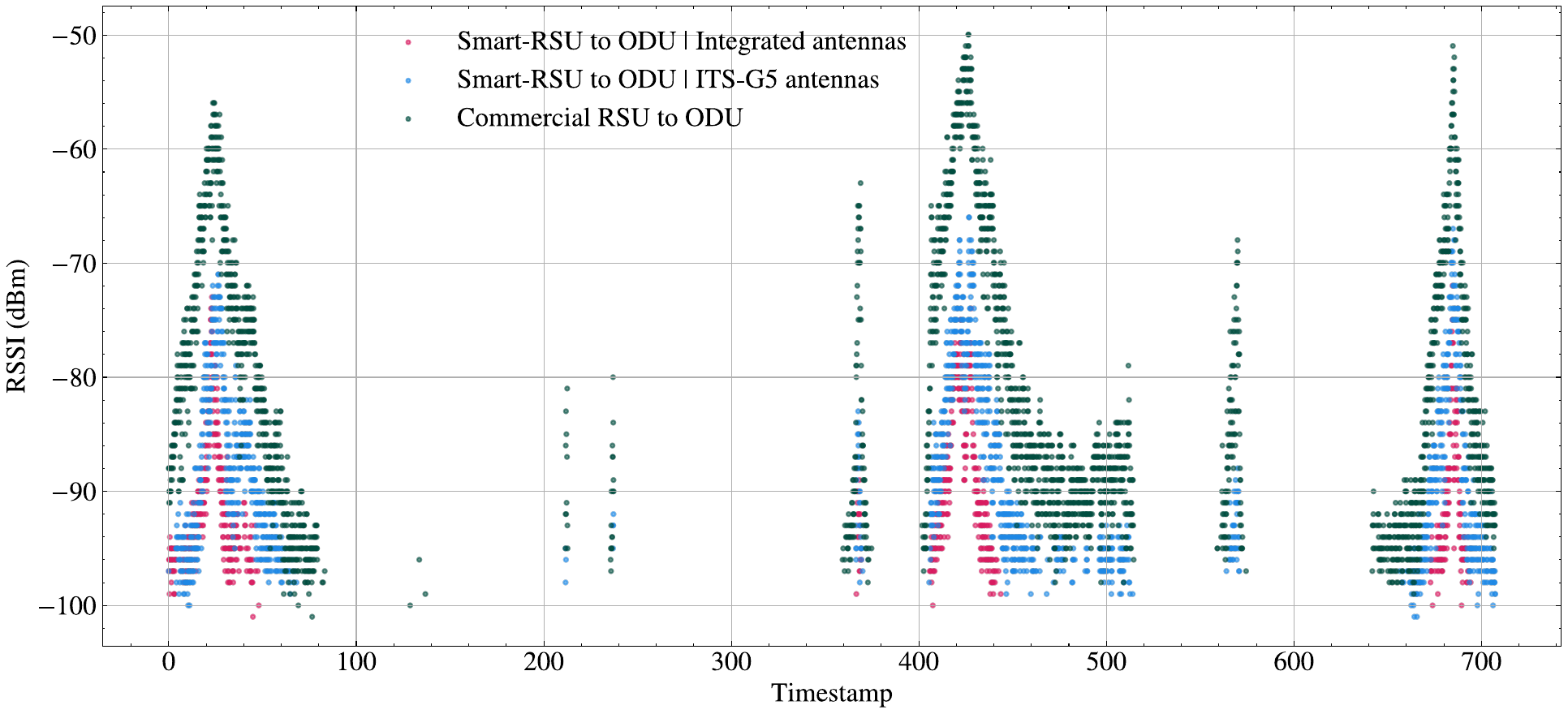}
\caption{Average RSSI received by the ODU over time from all RSUs during a representative test run.}
\label{fig:rsu_to_odu_rssi}
\end{figure}

\begin{figure}
    \centering
    \includegraphics[width=0.95\linewidth]{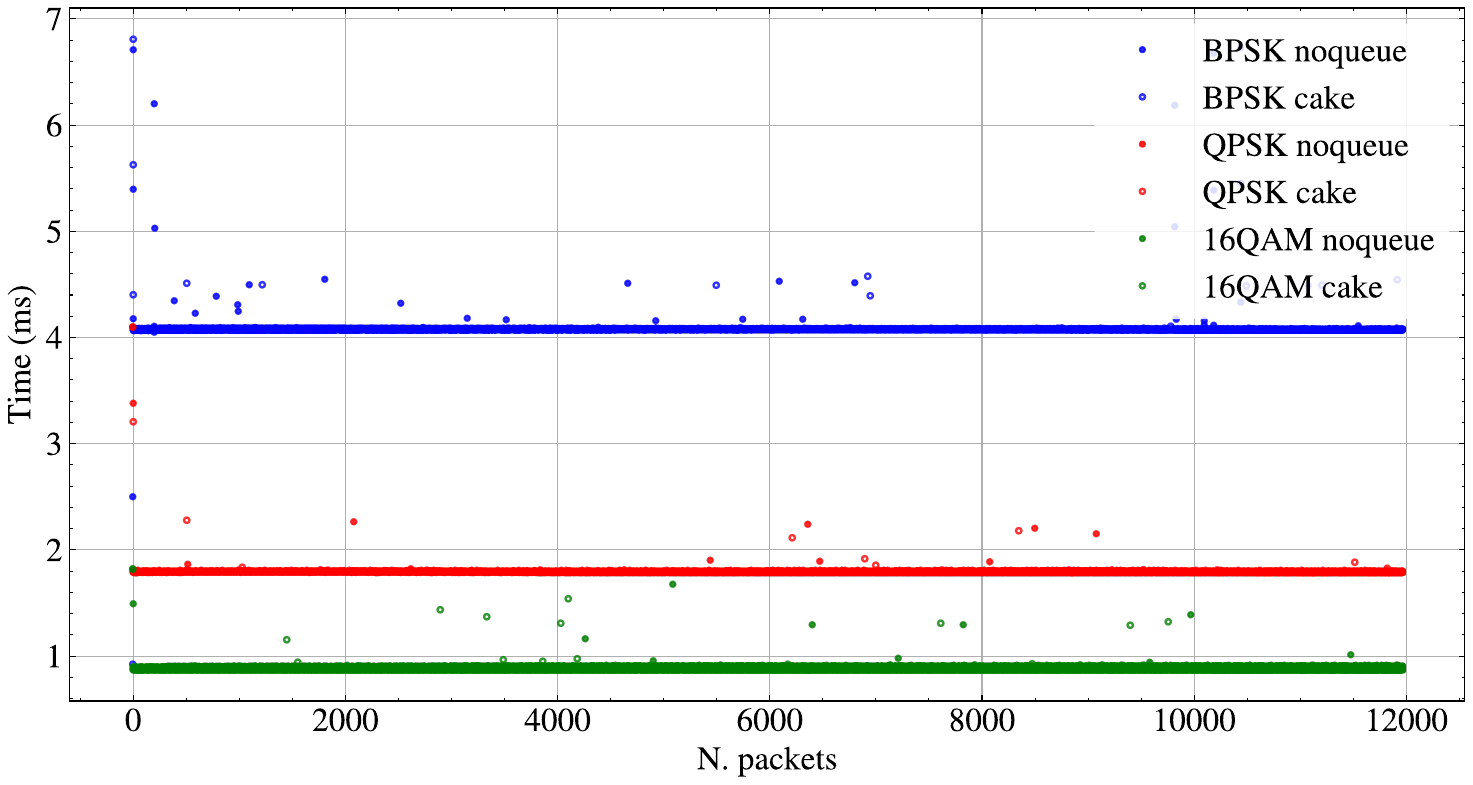}
\caption{Latency of a 2~Mbps UDP transmission as a function of the modulation scheme.}
\label{fig:udp_latency}
\end{figure}

This section presents the experimental results obtained from the measurement campaign, focusing on the impact of antenna configuration on V2X communication performance. In particular, we analyze received signal strength indicator (RSSI), packet loss, and latency in both uplink (OBU-to-infrastructure) and downlink (infrastructure-to-OBU) directions.

\subsection{OBU-to-Infrastructure Communication}

Fig.~\ref{fig:odu_to_rsu_rssi} shows the average RSSI measured at the SmartRSUs as a function of the distance between the vehicle (ODU) and the RSU pole. Since the SmartRSUs provide access to per-antenna measurements, the reported values correspond to the average RSSI across all available antennas to simplify visualization and comparison.

The results highlight a clear difference between the two antenna configurations. The SmartRSU equipped with external ITS-G5 antennas consistently achieves higher RSSI values compared to the integrated solution. Moreover, the coverage range is significantly extended: while the integrated antenna configuration provides reliable reception up to approximately $175~m$, the external antenna solution maintains connectivity up to approximately $300~m$, as shown in more detail in Fig.~\ref{fig:maprssi}. This behavior confirms that antenna placement and design directly affect signal propagation and reception quality, particularly in realistic environments where non-line-of-sight (NLOS) conditions are frequent. Fig.~\ref{fig:maploss} complements this analysis by reporting the spatial distribution of packet loss for the CAM messages transmitted by the ODU and received at the RSUs, highlighting the spatial heterogeneity of communication reliability. For clarity, only values below 50\% packet loss are shown, and results are aggregated over both time and space by averaging every 100 received messages.
This confirms that RSSI alone is not sufficient to characterize communication performance, especially in NLOS conditions.

The results again highlight the differences among the considered RSUs. The commercial RSU achieves the most extensive coverage, maintaining connectivity even in partially obstructed and non-line-of-sight regions. The SmartRSU equipped with external ITS-G5 antennas approaches similar performance along the main road, while showing increased sensitivity in lateral streets characterized by stronger NLOS conditions. In contrast, the SmartRSU with integrated antennas exhibits the highest packet loss and the most limited coverage area. These findings are consistent with the RSSI analysis and reinforce the importance of antenna design in real-world deployment scenarios.

\subsection{Infrastructure-to-OBU Communication}

Fig.~\ref{fig:rsu_to_odu_rssi} reports the average RSSI measured at the ODU over time during a representative test run. In this case, the ODU receives transmissions from all three RSUs, enabling a direct comparison between the two SmartRSUs and the commercial reference device.

It is worth noting that, while RSSI measurements at the infrastructure side are available only for the custom SmartRSUs, the ODU allows the extraction of RSSI values for all received packets, including those transmitted by the commercial RSU. This enables a complete comparison across all devices.

The results confirm the trends observed in the uplink analysis. The SmartRSU with external ITS-G5 antennas consistently outperforms the integrated solution, providing higher RSSI values over time. The commercial RSU exhibits a similar trend, but with an additional gain, likely due to optimized hardware design and antenna characteristics.

Overall, these results demonstrate that antenna configuration plays a key role in determining communication performance, affecting both coverage and signal quality. The comparison with the commercial RSU further highlights the gap between integrated solutions and more specialized antenna designs, while confirming the effectiveness of the proposed external antenna configuration.

\subsection{Latency Analysis under Static Transmission}

Fig.~\ref{fig:udp_latency} reports the latency measured during a controlled UDP transmission between the ODU and the SmartRSUs. The test is performed under static conditions, with the vehicle positioned near the RSU pole, using an IPv4 unicast flow generated with \texttt{iperf3} at a constant rate of $2~Mbps$.

For clarity, results are shown for a single SmartRSU, as both custom devices exhibit comparable latency behavior. The objective of this test is to validate the achievable transmission latency under stable channel conditions, corresponding to a low-resolution video stream typically required in remote driving applications. Thanks to full control of the V2X stack on both the OBU and RSU sides, IP-based unicast communication can be employed for this analysis, which is generally not available in commercial off-the-shelf devices.

The results show that latency strongly depends on the selected modulation scheme. In particular, BPSK yields an average latency of approximately $4~ms$, which decreases to below $2~ms$ with QPSK and to approximately $1~ms$ with 16QAM. These values confirm that low-latency transmission is achievable even with moderate data rates. Additionally, no significant variations are observed when changing the network interface's queue management policy. This indicates that, under the considered traffic conditions, latency is primarily determined by the physical layer configuration rather than by higher-layer buffering mechanisms.

Overall, these findings demonstrate that the proposed platform can support low-latency communication suitable for time-sensitive applications and enable fine-grained control and analysis of the networking stack.
These values are well within the latency requirements of real-time V2X applications.

\section{Conclusions}
\label{sec_conclusions}

This paper presented a measurement-based evaluation of custom SmartRSUs with different antenna configurations for V2X communication, deployed within the MASA living lab. Using a controlled yet realistic experimental setup, we performed a direct comparison among two in-house SmartRSUs and a commercial off-the-shelf RSU co-located on the same infrastructure.

The results highlight the strong impact of antenna design on communication performance. The SmartRSU equipped with external ITS-G5 antennas consistently outperforms the integrated solution in terms of RSSI, coverage, and packet loss, approaching the performance of the commercial RSU in several conditions. In contrast, the integrated antenna configuration, while more compact, exhibits reduced communication range and increased sensitivity to non-line-of-sight environments. In addition, latency measurements under controlled traffic conditions demonstrate that low-latency communication in the order of a few milliseconds is achievable with the proposed platform, confirming its suitability for time-sensitive applications such as remote driving. The ability to access and control the full V2X communication stack further enables detailed performance analysis, which is typically not possible with commercial devices.

Overall, this work confirms that SmartRSUs based on open and modular platforms can provide a flexible and effective alternative to proprietary solutions, while enabling reproducible experimentation and rapid prototyping of advanced V2X applications. At the same time, the results emphasize that antenna configuration remains a critical design factor that must be carefully considered in real-world deployments.

Future work will extend this analysis to more complex scenarios, including higher traffic densities, multi-RSU cooperation, and the integration of sensing and communication functionalities for infrastructure-assisted applications, reinforcing the role of smart infrastructure as a key enabler for next-generation cooperative and automated driving systems.

\balance
\bibliographystyle{IEEEtran}
\bibliography{bibliography}

\end{document}